\documentclass[12pt]{article}
\usepackage{amssymb}

\input{tcilatex}

\begin{document}

\bigskip\ 

\bigskip\ 

\begin{center}
\textbf{CHIROTOPE CONCEPT IN VARIOUS SCENARIOS OF PHYSICS}

\textbf{\ }

\textbf{\ }

\smallskip\ 

J. A. Nieto\footnote{%
nieto@uas.uasnet.mx}

\smallskip

\textit{Facultad de Ciencias F\'{\i}sico-Matem\'{a}ticas de la Universidad
Aut\'{o}noma}

\textit{de Sinaloa, 80010, Culiac\'{a}n Sinaloa, M\'{e}xico}

\bigskip\ 

\bigskip\ 

\textbf{Abstract}
\end{center}

We argue that the chirotope concept of oriented matroid theory may be found
in different scenarios of physics, including classical mechanics, quantum
mechanics, gauge field theory, $p$-branes formalism, two time physics and
Matrix theory. Our observations may motivate the interest of possible
applications of matroid theory in physics.

\bigskip\ 

\bigskip\ 

\bigskip\ 

\bigskip\ 

\bigskip\ 


Pacs numbers: 04.60.-m, 04.65.+e, 11.15.-q, 11.30.Ly

July, 2004

\newpage \bigskip

\noindent \textbf{1.- Introduction}

\smallskip\ 

Since Whitney's work [1], the concept of matroid has been of much interest
to a large number of mathematicians, specially those working in
combinatorial. Technically, this interest is perhaps due to the fact that
matroid theory [2] provides a generalization of both matrix theory and graph
theory. However, at some deeply level, it seems that matroid theory may
appear interesting to mathematicians, among other reasons, because its
duality properties. In fact, one of the attractive features of a matroid
theory is that every matroid has an associated dual matroid. This duality
characteristic refers to any individual matroid, but matroid theory states
stronger theorem at the level of axiom systems and their consequent
theorems, namely if there is an statement in the matroid theory that has
been proved true, then also its dual is true [3]. These duality propositions
play a so important role that matroid theory may even be called the duality
theory.

It turns out that at present the original formalism of matroid theory has
been generalized in different fronts, including biased matroids [4] and
greedoids [5]. However, it seems that one of the most natural extensions is
oriented matroid theory [6]. In turn, the matroid bundle structure [7]-[11]
emerges as a natural extension of oriented matroid theory. This final
extension provides with a very good example of the observation that two
fundamental mathematical subjects which have been developed independently,
are, sooner or later, fused in just one subject: in this case, fiber bundle
theory becomes fused with matroid theory leading to matroid bundle structure.

The central idea of the present work is to call the attention of the
physicists community about the possible importance that matroid theory may
have in different scenarios of physics. For this purpose in section \ 2 it
is developed a brief introduction of oriented matroid theory in such a way
that help us to prepare the mathematical tools which may facilitate its
connection with different scenarios of physics. In particular we introduce
the definition of an oriented matroid in terms of chirotopes (see Ref.[6]
section 3.5). Roughly speaking a chirotope is a completely antisymmetric
object that takes values in the set $\{-1,0,1\}$. It has been shown [12]
that the completely antisymmetric Levi-Civita symbol $\varepsilon
^{i_{1}...i_{d}}$ provides us with a particular example of a chirotope.
Motivated by this observation and considering that physicists are more or
less familiar with the symbol $\varepsilon ^{i_{1}...i_{d}}$ we develop a
brief introduction to oriented matroid theory by using the argument that the
chirotope concept is in fact a generalization of the symbol $\varepsilon
^{i_{1}...i_{d}}$. We hope that with such an introduction some physicists
become interested in the subject.

It is worth mentioning that the concept of matroid has already been
connected with Chern-Simons theory [13], string theory [14] and, $p$-branes
and Matrix theory [12]. Moreover, a proposed new theory called gravitoid
[15]-[16] has emerged from the connection between oriented matroid theory
and, gravity and supergravity. Except for the link between matroids and, $p$%
-branes and Matrix theory which are briefly reviewed here, all these
applications of the matroid concept are not approached in this work. Instead
here, we add new connections such as the identification of chirotopes with
the angular momentum in both classical and quantum mechanics scenarios. We
also remark the fundamental importance that chirotope concept may have in
two time physics [17] and, in electromagnetism and Yang-Mills physics.

In a sense, all these connections are similar to the identification of
tensors in different scenarios of physics. But , of course, although
interesting these identifications still appear more important the fact that
tensor analysis was eventually used as a the mathematical basis of a
fundamental theory: general relativity. The guide in this case was a new
symmetry provided by the equivalence principle, namely general covariance.
Therefore, the hope is that all these connections of matroids with different
concepts of physics may eventually help to identify a new fundamental theory
in which oriented matroid theory plays a basic role. But for this to be
possible we need a new symmetry as a guide. Our conjecture is that such a
fundamental theory is M-theory\ and that the needed guide symmetry is
duality. As, it is known M-theory [18]-[20] was suggested by various
dualities symmetries in string and $p$-brane theory. One of the interesting
aspects is that in oriented matroid theory duality is also of fundamental
importance as ordinary matroid theory (see Ref. [6] section 3.4). In fact,
there is also a theorem that establishes that every oriented matroid has and
associated dual oriented matroid. This is of vital importance for our
conjecture because if we write an action in terms of a given oriented
matroid we automatically assure an action for the dual oriented matroid and
as consequence the corresponding partition function must have a manifest
dual symmetry as seems to be required by M-theory.

By taking this observations as motivation in this article, we put special
emphasis in the chirotope concept identifying it in various scenarios of
physics. In section 2, it is introduced the concept of oriented matroid via
the chirotope concept. In section 3, it is made the identification of the
angular momentum in both classical and quantum mechanics with the chirotope
concept. In section 4, it is briefly review the connection between
chirotopes and $p$-branes. In section 5, we also briefly review the
connection between Matrix theory and matroids. In section 6, we made some
comments about the importance of the chirotope concept in two time physics.
Finally, in section 7 we make some final remarks explaining a possible
connection between the chirotope concept with electromagnetism and
Yang-Mills.

\bigskip\ 

\noindent \textbf{2.- Oriented matroid theory for physicists: a brief
introduction}

\smallskip\ 

The idea of this section is to give a brief introduction to the concept of
oriented matroid. But instead of following step by step the traditional
mathematical method presented in most teaching books (see [6] and Refs.
there in) of the subject we shall follow different route based essentially
in tensor analysis.

Let us start introducing the completely antisymmetric symbol

\begin{equation}
\varepsilon ^{i_{1}...i_{d}},  \tag{1}
\end{equation}%
which is, more or less, a familiar object in physics. (Here the indices $%
i_{1},...,i_{d}$ run from $1$ to $d$.) This is a rank-$d$ tensor which
values are $+1$ or $-1$ depending of even or odd permutations of%
\begin{equation}
\varepsilon ^{12...d},  \tag{2}
\end{equation}%
respectively. Moreover, $\varepsilon ^{i_{1}...i_{d}}$ takes the value $0$
unless $i_{1}...i_{d}$ are all different. In a more abstract and compact
form we can say that

\begin{equation}
\varepsilon ^{i_{1}...i_{d}}\in \{-1,0,1\}.  \tag{3}
\end{equation}%
An important property of $\varepsilon ^{i_{1}...i_{d}}$ is that has exactly
the same number of indices as the dimension $d$ of the space.

Another crucial property of the symbol $\varepsilon ^{i_{1}...i_{d}}$ is
that the product $\varepsilon ^{i_{1}...i_{d}}\varepsilon ^{j_{1}...j_{d}}$
can be written in terms of a product of the Kronecker deltas $\delta
^{ij}=diag(1,...,1).$ Specifically, we have 
\begin{equation}
\varepsilon ^{i_{1}...i_{d}}\varepsilon ^{j_{1}...j_{d}}=\delta
^{i_{1}...i_{d},j_{1}...j_{d}},  \tag{4}
\end{equation}%
where $\delta ^{i_{1}...i_{d},j_{1}...j_{d}}$ is the so called delta
generalized symbol;

\begin{equation}
\delta ^{i_{1}...i_{d}j_{1}...j_{d}}=\left\{ 
\begin{array}{c}
+1\text{ if }i_{1}...i_{d}\text{ is an even permutation of }j_{1}...j_{d},
\\ 
-1\text{ if }i_{1}...i_{d}\text{ is an odd permutation of }j_{1}...j_{d}, \\ 
\multicolumn{1}{l}{\text{ }0\text{ otherwise}.}%
\end{array}%
\right.  \tag{5}
\end{equation}%
An example may help to understand the $\delta ^{i_{1}...i_{d},j_{1}...j_{d}}$
symbol. Assume that $d$ is equal $2.$ Then we have $\varepsilon
^{i_{1}i_{2}} $ and

\begin{equation}
\varepsilon ^{i_{1}i_{2}}\varepsilon ^{j_{1}j_{2}}=\delta
^{i_{1}i_{2},j_{1}j_{2}}=\delta ^{i_{1},j_{1}}\delta ^{i_{2},j_{2}}-\delta
^{i_{1}j_{2}}\delta ^{i_{2},j_{1}}.  \tag{6}
\end{equation}

From (4) it follows the antisymmetrized square bracket property

\begin{equation}
\varepsilon ^{i_{1}...[i_{d}}\varepsilon ^{j_{1}...j_{d}]}\equiv 0.  \tag{7}
\end{equation}%
We recall that for any tensor $V^{i_{1}i_{2}.i_{3}}$ the object $%
V^{[i_{1}i_{2}.i_{3}]}$ is defined by

\[
V^{[i_{1}i_{2}.i_{3}]}=\frac{1}{3!}%
(V^{i_{1}i_{2}.i_{3}}+V^{i_{2}i_{3}.i_{1}}+V^{i_{3}i_{1}.i_{2}}-V^{i_{2}i_{1}.i_{3}}-V^{i_{1}i_{3}.i_{2}}-V^{i_{3}i_{2}.i_{1}}), 
\]%
with obvious generalization to any dimension. The result (7) comes from the
fact that any completely antisymmetric tensor with more than $d$ indices
must vanish. Indeed it can be shown that any completely antisymmetric tensor 
$F^{i_{1}...i_{r}}$ with $r>d$ must vanish, while if $r=d,$ $%
F^{i_{1}...i_{n}}$ must be proportional to $\varepsilon ^{i_{1}...i_{d}}$.
In other words, up to a factor the symbol $\varepsilon ^{i_{1}...i_{d}}$ is
the largest completely antisymmetric tensor that one can have in $d$
dimensions.

Now, we would like to relate the symbol $\varepsilon ^{i_{1}...i_{d}}$ with
the chirotope concept of oriented matroid theory. For this purpose we ask
ourselves whether it is possible having the analogue of the symbol $%
\varepsilon ^{i_{1}...i_{d}}$ for $r<d.$ There is not any problem for having
completely antisymmetric tensors $F^{i_{1}...i_{r}}$ for $r<d,$ why then not
to consider the analogue of $\varepsilon ^{i_{1}...i_{d}}$ for $r<d$? Let us
denote by $\sigma ^{i_{1}...i_{r}},$ with $r<d,$ this assumed analogue of $%
\varepsilon ^{i_{1}...i_{d}}$. What properties should we require for the
object $\sigma ^{i_{1}...i_{r}}$? According to our above discussion one may
say that $\varepsilon ^{i_{1}...i_{d}}$ is determined by the properties (3)
and (7). Therefore, we require exactly similar properties for $\sigma
^{i_{1}...i_{r}},$ namely $\sigma ^{i_{1}...i_{r}}$ is a completely
antisymmetric under interchange of any pair of indices and satisfy the two
conditions,

\begin{equation}
\sigma ^{i_{1}...i_{r}}\in \{-1,0,1\}  \tag{8}
\end{equation}%
and

\begin{equation}
\sigma ^{i_{1}...[i_{r}}\sigma ^{j_{1}...j_{r}]}\equiv 0.  \tag{9}
\end{equation}

A solution for (9) is provided by

\begin{equation}
\Sigma ^{i_{1}...i_{r}}=\varepsilon
^{a_{1}...a_{r}}v_{a_{1}}^{i_{1}}...v_{a_{r}}^{i_{r}},  \tag{10}
\end{equation}%
where $v_{a}^{i}$ is any $r\times d$ matrix over some field $F$. Other way
to write (10) is

\begin{equation}
\Sigma ^{i_{1}...i_{r}}=\det (\mathbf{v}^{i_{1}}...\mathbf{v}^{i_{r}}). 
\tag{11}
\end{equation}%
One may prove that (10) implies (9) as follows. Assuming (10) we get

\begin{equation}
\begin{array}{ccc}
\Sigma ^{i_{1}...[i_{r}}\Sigma ^{j_{1}...j_{r}]} & = & \varepsilon
^{a_{1}...a_{r}}\varepsilon
^{b_{1}...b_{r}}v_{a_{1}}^{i_{1}}...v_{a_{r}}^{[i_{r}}v_{b_{1}}^{j_{1}}...v_{b_{r}}^{j_{r}]}
\\ 
&  &  \\ 
& = & \varepsilon ^{a_{1}...[a_{r}}\varepsilon
^{b_{1}...b_{r}]}v_{a_{1}}^{i_{1}}...v_{a_{r}}^{i_{r}}v_{b_{1}}^{j_{1}}...v_{b_{r}}^{j_{r}}.%
\end{array}
\tag{12}
\end{equation}%
But from (7) we know that

\begin{equation}
\varepsilon ^{a_{1}...[a_{r}}\varepsilon ^{b_{1}...b_{r}]}=0,  \tag{13}
\end{equation}%
and therefore we find

\begin{equation}
\Sigma ^{i_{1}...[i_{r}}\Sigma ^{j_{1}...j_{r}]}=0,  \tag{14}
\end{equation}%
as required.

Since $\det (\mathbf{v}^{i_{1}}...\mathbf{v}^{i_{r}})$ can be positive,
negative or zero we may have a tensor $\sigma ^{i_{1}...i_{r}}$ satisfying
both (3) and (7) by setting

\begin{equation}
\sigma ^{i_{1}...i_{r}}=sign\Sigma ^{i_{1}...i_{r}}.  \tag{15}
\end{equation}%
Observe that if $r=d$ and $v_{a}^{i}$ is the identity then $\sigma
^{i_{1}...i_{d}}=\varepsilon ^{i_{1}...i_{d}}$. Therefore the tensor $\sigma
^{i_{1}...i_{r}}$ is a more general object than $\varepsilon
^{i_{1}...i_{d}} $.

Let us now analyze our results from other perspective. First, instead of
saying that the indices $i_{1}...i_{d}$ run from $1$ to $d$ we shall say
that the indices $i_{1}...i_{d}$ take values in the set $E=\{1,...,d\}.$In
other words we set

\begin{equation}
i_{1}...i_{d}\in \{1,...,d\}.  \tag{16}
\end{equation}%
Now, suppose that to each element of $E$ we associate a $r-$dimensional
vector $\mathbf{v}$. In other word, we assume the map

\begin{equation}
i\rightarrow \mathbf{v(}i\mathbf{)}\equiv \mathbf{v}^{i}.  \tag{17}
\end{equation}%
We shall write the vector $\mathbf{v}^{i}$ as $v_{a}^{i}$, with $a\in
\{1,...,r\}$. With this notation the map (17) becomes

\begin{equation}
i\rightarrow v_{a}^{i}.  \tag{18}
\end{equation}

Let us try to understand the expression (10) in terms of a family-set. First
note that because the symbol $\varepsilon ^{a_{1}...a_{r}}$ makes sense only
in $r-$dimensions the indices $i_{1}...i_{r}$ combination in $\Sigma
^{i_{1}...i_{r}}$ corresponds to $r-$elements subsets of $E=\{1,...,d\}.$
This motive to define the family $\mathcal{B}$ of all possible $r-$elements
subsets of $E.$

An example may help to understand our observations. Consider the object

\begin{equation}
\Sigma ^{ij}.  \tag{19}
\end{equation}%
We establish that

\begin{equation}
i,j\in E=\{1,2,3\}.  \tag{20}
\end{equation}

Assume that

\begin{equation}
\Sigma ^{ij}=-\Sigma ^{ji},  \tag{21}
\end{equation}%
that is $\Sigma ^{ij}$ is an antisymmetric second rank tensor \ This means
that the only nonvanishing components of $\Sigma ^{ij}$ are $\Sigma
^{12},\Sigma ^{13}$ and $\Sigma ^{23}.$ From these nonvanishing components
of $\Sigma ^{ij}$ we may propose the family-set

\begin{equation}
\mathcal{B}=\{\{1,2\},\{1,3\},\{2,3\}\}.  \tag{22}
\end{equation}%
Further, suppose we associate to each value of $i$ a two dimensional vector $%
\mathbf{v}(i)$. This means that the set $E$ can be written as

\begin{equation}
E=\{\mathbf{v}(1),\mathbf{v}(2),\mathbf{v}(3)\}.  \tag{23}
\end{equation}%
This process can be summarizing by means of the transformation

\begin{equation}
i\rightarrow v_{a}^{i},  \tag{24}
\end{equation}%
with $a\in \{1,2\}$. We can connect $v_{a}^{i}$ with an explicit form of $%
\Sigma ^{ij}$ if we write

\begin{equation}
\Sigma ^{ij}=\varepsilon ^{ab}v_{a}^{i}v_{b}^{j}.  \tag{25}
\end{equation}

The previous considerations proof the possible existence of an object such
as $\sigma ^{i_{1}...i_{r}}$. In the process of proposing the object $\sigma
^{i_{1}...i_{r}}$ we have introduced the set $E$ and the $r-$element subsets 
$\mathcal{B}$. It turns out that the pair $(E,\mathcal{B})$ plays an
essential role in the definition of a matroid. But before we formally define
a matroid, we would like to make one further observation. For this purpose
we first notice that (9) implies%
\begin{equation}
\sigma ^{i_{1}...i_{r}}\sigma ^{j_{1}...j_{r}}\equiv \sum_{a=1}^{r}\sigma
^{j_{a}i_{2}...i_{r}}\sigma ^{j_{1}..j_{a-1}.i_{1}j_{a+1}...j_{r}}.  \tag{26}
\end{equation}%
Therefore, if $\sigma ^{i_{1}...i_{r}}\sigma ^{j_{1}...j_{r}}\neq 0$ the
expression (26) means that there exist an $a\in \{1,2,...,r\}$ such that

\begin{equation}
\sigma ^{i_{1}...i_{r}}\sigma ^{j_{1}...j_{r}}\equiv \sigma
^{j_{a}i_{2}...i_{r}}\sigma ^{j_{1}..j_{a-1}.i_{1}j_{a+1}...j_{r}}.  \tag{27}
\end{equation}%
This prove that (9) implies (27) but the converse is not true. Therefore,
the expression (27) defines an object that it is more general than one
determined by (9). Let us denote this more general object by $\chi
^{i_{1}...i_{r}}$. We are ready to formally define an oriented matroid (see
Ref. [6] section 3.5):

Let $r\geq 1$ be an integer, and let $E$ be a finite set (ground set). An
oriented matroid $\mathcal{M}$ of rank $r$ is the pair $(E,\chi )$ where $%
\chi $ is a mapping (called chirotope) $\chi :E\rightarrow \{-1,0,1\}$ which
satisfies the following three properties:

1) $\chi $ is not identically zero

2) $\chi $ is completely antisymmetric.

3) for all $i_{1},...,i_{r},j_{1},...,j_{r}\in E$ such that

\begin{equation}
\chi ^{i_{1}...i_{r}}\chi ^{j_{1}...j_{r}}\neq 0.  \tag{28}
\end{equation}%
There exist and $a$ such that

\begin{equation}
\chi ^{i_{1}...i_{r}}\chi ^{j_{1}...j_{r}}=\chi ^{j_{a}i_{2}...i_{r}}\chi
^{j_{1}..j_{a-1}.i_{1}j_{a+1}...j_{r}}.  \tag{29}
\end{equation}

Let $\mathcal{B}$ be the set of $r-$elements subsets of $E$ such that

\begin{equation}
\chi ^{i_{1}...i_{r}}\neq 0,  \tag{30}
\end{equation}%
for $i_{1},...,i_{r}\in E$. Then (29) implies that if $i_{a}\in B$ there
exist $j_{a}\in B^{\prime }\in $ $\mathcal{B}$ such that $(B-i_{a})\cup
j_{a}\in \mathcal{B}$. This important property of the elements of $\mathcal{B%
}$ defines an ordinary matroid on $E$ (see Ref. [2] section 1.2).

Formally, a matroid $M$ is a pair $(E,\mathcal{B})$, where $E$ is a
non-empty finite set and $\mathcal{B}$ is a non-empty collection of subsets
of $E$ (called bases) satisfying the following properties:

$(\mathcal{B}$ $\mathit{i)}$\textit{\ }no basis properly contains another
basis;

$(\mathcal{B}$ $\mathit{ii)}$ if $B_{1}$ and $B_{2}$ are bases and if $b$ is
any element of $B_{1},$ then there is an element $g$ of $B_{2}$ with the
property that $(B_{1}-\{b\})\cup \{g\}$ is also a basis.

$M$ is called the underlying matroid of $\mathcal{M}$. According to our
considerations every oriented matroid $\mathcal{M}$ has an associated
underlying matroid $M$. However the converse is not true, that is, not every
ordinary matroid $M$ has an associated oriented matroid $\mathcal{M}$. In a
sense this can be understood observing that (29) not necessarily implies
condition (9). In other words, the condition (29) is less restrictive than
(9). It is said that an ordinary matroid $M$ is orientable if there is an
oriented matroid $\mathcal{M}$ with an underlying matroid $M$. There are
many examples of non-oriented matroids, perhaps one of the most interesting
is the so called Fano matroid $F_{7}$ (see Ref. [6] section 6.6). This is a
matroid defined on the ground set

\[
E=\{1,2,3,4,5,6,7\}, 
\]%
whose bases are all those subsets of $E$ with three elements except $%
f_{1}=\{1,2,3\},$ $f_{2}=\{5,1,6\},$ $f_{3}=\{6,4,2\},$ $f_{4}=\{4,3,5\},$ $%
f_{5}=\{4,7,1\},$ $f_{6}=\{6,7,3\}$ and $f_{7}=\{5,7,2\}$. This matroid is
realizable over a binary field and is the only minimal irregular matroid.
Moreover, it has been shown [13]-[16] that $F_{7}$ is connected with
octionions and therefore with supergravity. However, it appears intriguing
that in spite these interesting properties of $F_{7}$ this matroid is not
orientable.

It can be shown that all bases have the same number of elements. The number
of elements of a basis is called rank and we shall denote it by $r.$ Thus,
the rank of an oriented matroid is the rank of its underlying matroid.

One of the simplest, but perhaps one of the most important, ordinary
matroids is the so call it uniform matroid denoted as $U_{r,d}$ and defined
by the pair $(E,\mathcal{B)}$, where $E=\{1,...,d\}$ and $\mathcal{B}$ is
the collection of $r-$element subsets of $E$, with $r\leq d.$

With these definitions at hand we can now return to the object $\varepsilon
^{i_{1}...i_{d}}$ and reanalyze it in terms of the oriented matroid concept.
The tensor $\varepsilon ^{i_{1}...i_{d}}$ has an associated set $%
E=\{1,2,...,d\}$. It is not difficult to see that in this case $\mathcal{B}$
is given by $\{\{1,2,...,d\}\}.$ This means that the only basis in $\mathcal{%
B}$ is $E$ itself. Further since $\varepsilon ^{i_{1}...i_{d}}$ satisfies
the property (7) must also satisfy the condition (29) and therefore we have
discovered that $\varepsilon ^{i_{1}...i_{d}}$ is a chirotope, with
underlying matroid $U_{d,d}$. Thus, our original question whether is it
possible to have the analogue of the symbol $\varepsilon ^{i_{1}...i_{d}}$
for $r<d$ is equivalent to ask wether there exist chirotopes for $r<d$ and
oriented matroid theory give us a positive answer. An object $\chi
^{i_{1}...i_{r}}$ satisfying the definition of oriented matroid is a
chirotope that, in fact, generalize the symbol $\varepsilon ^{i_{1}...i_{d}}$%
.

A realization of $\mathcal{M}$ is a mapping $\mathbf{v}:E\rightarrow R^{r}$
such that

\begin{equation}
\chi ^{i_{1}...i_{r}}\rightarrow \sigma ^{i_{1}...i_{r}}=sign\Sigma
^{i_{1}...i_{r}},  \tag{30}
\end{equation}%
for all $i_{1},...,i_{r}\in E$. Here, $\Sigma ^{i_{1}...i_{r}}$ is given in
(10). By convenience we shall call the symbol $\Sigma ^{i_{1}...i_{r}}$ 
\textit{prechirotope}.

Realizability is a very important subject in oriented matroid theory and
deserves to be discussed in some detail. However, in this paper we are more
interested in a rough introduction to the subject and for that reason we
refer to the interested redear to the Chapter 8 of reference [6] where a
whole discussion of the subject is given. Nevertheless, we need to make some
important remarks. First of all, it turns out that not all oriented matroids
are realizable. In fact, it has been shown that the smallest non-ralizable
uniform oriented matroids have the $(r,d)$-parameters $(3,9)$ and $(4,8).$
It is worth mentioning that given a uniform matroid $U_{r,d}$ the
orientability is not unique. For instance, there are precisely 2628
(reorientations classes of) uniform $r=4$ oriented matroids with $d=8$.
Further, precisely 24 of these oriented matroids are non-realizables.

A rank preserving weak map concept is another important notion in oriented
matroid theory. This is a map between two oriented matroids $\mathcal{M}_{1}$
and $\mathcal{M}_{2}$ on the same ground set $E$ and $r_{1}=r_{2}$ with the
property that every basis of $\mathcal{M}_{2}$ is a basis of $\mathcal{M}%
_{1}.$ There is an important theorem that establishes that every oriented
matroid is the weak map image of uniform oriented matroid of the same rank.

Finally, we should mention that there is a close connection between
Grassmann algebra and chirotopes. To understand this connection let us
denote by $\wedge _{r}R^{n}$ the $(_{r}^{n})$-dimensional real vector space
of alternating $r$-forms on $R^{n}$. An element $\mathbf{\Sigma }$ in $%
\wedge _{r}R^{n}$ is said to be decomposable if

\begin{equation}
\mathbf{\Sigma }=\mathbf{v}_{1}\wedge \mathbf{v}_{2}\wedge ...\wedge .%
\mathbf{v}_{r},  \tag{31}
\end{equation}%
for some $\mathbf{v}_{1},\mathbf{v}_{2},...,.\mathbf{v}_{r}\in R^{n}$. It is
not difficult to see that (31) can be written as

\begin{equation}
\mathbf{\Sigma }=\frac{1}{r!}\Sigma ^{i_{1}...i_{r}}e_{i_{1}}\wedge
e_{i_{2}}\wedge ...\wedge e_{i_{r}},  \tag{32}
\end{equation}%
where $e_{i_{1}},e_{i_{2}},...,e_{i_{r}}$ are one form bases in $R^{n}$ and $%
\Sigma ^{i_{1}...i_{r}}$ is given in (10). This shows that the prechirotope $%
\Sigma ^{i_{1}...i_{r}}$ can be identified with an alternating decomposable $%
r$-forms. It is known that the projective variety of decomposable forms is
isomorphic to the Grassmann variety of $r$-dimensional linear subspaces in $%
R^{n}$. In turn, the Grassmann variety is the classifying space for vector
bundle structures. Perhaps, related observations motivate to MacPherson [7]
to develop the combinatorial differential manifold concept which was the
predecessor of the matroid bundle concept [7]-[11]. This is a differentiable
manifold in which at each point it is attached an oriented matroid as a
fiber.

It is appropriate to briefly comment about the origins of chirotope concept.
It seems that the concept of chirotope appears for the first time in 1965 in
a paper by Novoa [21] under the name "n-ordered sets and order
completeness". The term chirotope was used by Dress [22] in connection with
certain chirality structure in organic chemistry. Bokowski and Shemer [23]
applies the chirotope concept in relation with the Steinitz problem.
Finally, Las Verganas [24] used the chirotope concept to construct an
alternative definition of oriented matroid.

Now, the symbol $\varepsilon ^{i_{1}...i_{d}}$ is very much used in
different context of physics, including supergravity and $p$-branes.
Therefore the question arises whether the chirotope symbol $\chi
^{i_{1}...i_{r}}$ may have similar importance in different scenarios of
physics. In the next sections we shall make the observation that the symbol $%
\Sigma ^{i_{1}...i_{r}}$ is already used in different scenarios of physics,
but apparently it has not been recognized as a chirotope.

\bigskip\ 

\noindent \textbf{3.- Chirotopes in classical and quantum mechanics}

\smallskip\ 

It is well known that the angular momentum $\bar{L}$ in $3$-dimensional
space is one of the most basic concepts in classical mechanics.
Traditionally $\bar{L}$ is defined by

\begin{equation}
\bar{L}=\bar{r}\times \bar{p}.  \tag{33}
\end{equation}%
In tensor notation this expression can be written as

\begin{equation}
L^{i}=\varepsilon ^{ijk}x_{j}p_{k}.  \tag{34}
\end{equation}%
We observe the presence of the symbol $\varepsilon ^{ijk}$ which is a
chirotope. In fact, this $\varepsilon -$symbol appears in any cross product $%
\bar{A}\times \bar{B}$ for any two vectors $\bar{A}$ and $\bar{B}$ in 3
dimensions$.$ We still have a deeper connection between $\bar{L}$ and
matroids. First, we observe that the formula (34) can also be written as

\begin{equation}
L^{i}=\frac{1}{2}\varepsilon ^{ijk}L_{jk},  \tag{35}
\end{equation}%
where

\begin{equation}
L^{ij}=x^{i}p^{j}-x^{j}p^{i}.  \tag{36}
\end{equation}%
Of course, $L^{i}$ and $L^{ij}$ have the same information.

Let us redefine $x^{i}$ and $p^{j}$ in the form

\begin{equation}
\begin{array}{c}
v_{1}^{i}\equiv x^{i} \\ 
\\ 
v_{2}^{i}\equiv p^{i}.%
\end{array}
\tag{37}
\end{equation}%
Using this notation the expression (36) becomes

\begin{equation}
L^{ij}=\varepsilon ^{ab}v_{a}^{i}v_{b}^{j},  \tag{38}
\end{equation}%
where the indices $a$ and $b$ take values in the set $\{1,2\}$. If we
compare (38) with (10), we recognize in (38) the form of a rank$-2$
prechirotope. This means that the angular momentum itself is a prechirotope.
For a possible generalization to any dimension, the form (38) of the angular
momentum appears more appropriate than the form (35). Thus, our conclusion
that the angular momentum is a prechirotope applies to any dimension, not
just 3-dimensions.

The classical Poisson brackets associated to $L^{ij}$ is

\begin{equation}
\{L^{ij},L^{kl}\}=\delta ^{ik}L^{jl}-\delta ^{il}L^{jk}+\delta ^{jl}\delta
L^{ik}-\delta ^{jk}\delta L^{il}.  \tag{39}
\end{equation}%
One of the traditional mechanism for going from classical mechanics to
quantum mechanics is described by the prescription

\begin{equation}
\{A,B\}\rightarrow \frac{1}{i}[\hat{A},\hat{B}],  \tag{40}
\end{equation}%
for any two canonical variables $A$ and $B.$ Therefore, at the quantum level
the expression (39) becomes

\begin{equation}
\lbrack \hat{L}^{ij},\hat{L}^{kl}]=i(\delta ^{ik}\hat{L}^{jl}-\delta ^{il}%
\hat{L}^{jk}+\delta ^{jl}\hat{L}^{ik}-\delta ^{jk}\hat{L}^{il}).  \tag{41}
\end{equation}%
It is well known the importance of this expression in both the eigenvalues
determination and the group analyses of a quantum system. Therefore, the
prechirotope property of $L^{ij}$ goes over at the quantum level.

\bigskip\ 

\noindent \textbf{4.- Chirotopes and }$p$\textbf{-branes}

\smallskip\ 

Consider the action

\begin{equation}
S=\frac{1}{2}\int d^{p+1}\xi (\gamma ^{-1}\gamma ^{\mu _{1}...\mu
_{p+1}}\gamma _{\mu _{1}...\mu _{p+1}}-\gamma T_{p}^{2}),  \tag{42}
\end{equation}%
where%
\begin{equation}
\gamma ^{\mu _{1}...\mu _{p+1}}=\varepsilon ^{a_{1}...a_{p+1}}V_{a_{1}}^{\mu
_{1}}(\xi )...V_{a_{p+1}}^{\mu _{p+1}}(\xi ),  \tag{43}
\end{equation}%
with 
\begin{equation}
V_{a}^{\mu }(\xi )=\partial _{a}x^{\mu }(\xi ).  \tag{44}
\end{equation}%
Here $\gamma $ is a lagrange multiplier and $T_{p}$ is a constant measuring
the inertial of the system. It turns out that the action (42) is equivalent
to the Nambu-Goto type action for $p$-branes (see [12] and Refs there in).
One of the important aspects of (42) is that makes sense to set $T_{p}=0.$
In such case, (42) is reduced to the Schild type null $p$-brane action
[26]-[27].

From (43) we observe that, except for its locality, $\gamma ^{\mu _{1}...\mu
_{p+1}}$ has the same form as a prechirotope. The local property of $\gamma
^{\mu _{1}...\mu _{p+1}}$ can be achieved by means of the matroid bundle
concept. The key idea in matroid bundle is to replace tangent spaces in a
differential manifold by oriented matroids. This is achieved by considering
the linear map $f_{\xi }:\shortparallel star\Delta \shortparallel
\rightarrow U\subset T_{\eta (\xi )}$ such that $f_{\xi }(\xi )=0$, where $%
\shortparallel \Delta \shortparallel $ is the minimal simplex of $%
\shortparallel X\shortparallel $ containing $\xi \in X,$ where $X$ is a
simplicial complex associated to a differential manifold. Then, $f_{\xi
}\shortparallel (star\Delta )^{0}\shortparallel $, where $(star\Delta )^{0}$
are the $0$-simplices of $star\Delta ,$ is a configuration of vectors in $%
T_{\eta (\xi )}$ defining an oriented matroid $\mathcal{M}(\xi )$. One
should expect that the function $f_{\xi }$ induces a map

\begin{equation}
\Sigma ^{\mu _{1}...\mu _{r}}\rightarrow \gamma ^{\mu _{1}...\mu _{p+1}}(\xi
),  \tag{45}
\end{equation}%
where we consider that the rank $r$ of $\mathcal{M}(\xi )$ is $r=p+1$.
Observe that the formula (45) means that the function $f_{\xi }$ also
induces the map $v_{a}^{\mu }$ $\rightarrow V_{a}^{\mu }(\xi ).$

Our last task is to establish the expression (44). Consider the expression

\begin{equation}
F_{ab}^{\mu }=\partial _{a}V_{b}^{\mu }(\xi )-\partial _{b}V_{a}^{\nu }(\xi
).  \tag{46}
\end{equation}%
Thus, if the equation $F_{ab}^{\mu }=0$ is implemented in (42) as a
constraint then we get the solution $V_{a}^{\mu }(\xi )=\frac{\partial
x^{\mu }}{\partial \xi ^{a}},$ where $x^{\mu }$ is, in this context, a gauge
function. In this case, one says that $v_{a}^{\mu }(\xi )$ is a pure gauge.
Of course, $F_{ab}^{\mu }$ and $V_{b}^{\mu }(\xi )$ can be interpreted as
field strength and abelian gauge potential, respectively.

\bigskip\ 

\noindent \textbf{5.- Chirotopes and Matrix theory}

\smallskip\ 

Some years ago Yoneya [28] showed that it is possible to construct Matrix
theory the Schild type action for strings. The key idea in the Yoneya's work
is to consider the Poisson bracket structure

\begin{equation}
\{x^{\mu },x^{\nu }\}=\frac{1}{\xi }\gamma ^{\mu \nu },  \tag{47}
\end{equation}%
where $\xi $ is an auxiliary field. This identification suggests to replace
the Poisson structure by coordinate operators

\begin{equation}
\{x^{\mu },x^{\nu }\}\rightarrow \frac{1}{i}[\hat{x}^{\mu },\hat{x}^{\nu }].
\tag{48}
\end{equation}%
The next step is to quantize the constraint

\begin{equation}
-\frac{1}{\xi ^{2}}\gamma ^{\mu \nu }\gamma _{\mu \nu }=T_{p}^{2},  \tag{49}
\end{equation}%
which can be derived from (42) by setting $p=1$. According to (47), (48) and
(49) one gets

\begin{equation}
([\hat{x}^{\mu },\hat{x}^{\nu }])^{2}=T_{p}^{2}I,  \tag{50}
\end{equation}%
where $I$ is the identity operator. It turns out that the constraint (50)
plays an essential role in Matrix theory. Extending the Yoneya's idea for
strings, Oda [29] (see also [30]-[31]) has shown that it is also possible to
construct a Matrix model of M-theory from a Schild-type action for
membranes. It is clear from our previous analysis of identifying the
quantity $\gamma ^{\mu \nu }$ with a prechirotope of a given chirotope $\chi
^{\mu \nu }$ that these developments of Matrix theory can be linked with the
oriented matroid theory.

\bigskip\ 

\noindent \textbf{6.- Chirotopes and two time physics}

\smallskip\ 

Consider the first order lagrangian [17]

\begin{equation}
L=\frac{1}{2}\varepsilon ^{ab}\dot{v}_{a}^{\mu }v_{b}^{\nu }\eta _{\mu \nu
}-H(v_{a}^{\mu }),  \tag{51}
\end{equation}%
where $\eta _{\mu \nu }$ is a flat metric whose signature it will be
determined below. Up to total derivative this lagrangian is equivalent to
the first order lagrangian

\begin{equation}
L=\dot{x}^{\mu }p_{\mu }-H(x,p),  \tag{52}
\end{equation}%
where

\begin{equation}
\begin{array}{c}
x^{\mu }=v_{1}^{\mu }, \\ 
\\ 
p^{\mu }=v_{2}^{\mu }.%
\end{array}
\tag{53}
\end{equation}

Typically one chooses $H$ as $H=\lambda (p^{\mu }p_{\mu }+m^{2})$. For the
massless case we have

\begin{equation}
H=\lambda (p^{\mu }p_{\mu }).  \tag{54}
\end{equation}%
From the point of view of the lagrangian (51) in terms of the coordinates $%
v_{a}^{\mu }$ this choice is not good enough since the $SL(2,R)-$symmetry in
the first term of (51) is lost. It turns out that the simplest possible
choice for $H$ which maintains the symmetry $SL(2,R)$ is 
\begin{equation}
H=\frac{1}{2}\lambda ^{ab}v_{a}^{\mu }v_{b}^{\nu }\eta _{\mu \nu },  \tag{55}
\end{equation}%
where $\lambda ^{ab}$ is a lagrange multipliers. Arbitrary variations of $%
\lambda ^{ab}$ lead to the constraint $v_{a}^{\mu }v_{b}^{\nu }\eta _{\mu
\nu }=0$ which means that

\begin{equation}
p^{\mu }p_{\mu }=0,  \tag{56}
\end{equation}

\begin{equation}
p^{\mu }x_{\mu }=0  \tag{57}
\end{equation}%
and

\begin{equation}
x^{\mu }x_{\mu }=0.  \tag{58}
\end{equation}%
The key point in two time physics comes from the observation that if $\eta
_{\mu \nu }$ corresponds to just one time, that is, if $\eta _{\mu \nu }$
has the signature $\eta _{\mu \nu }=diag(-1,1,...,1)$ then from (56)-(58) it
follows that $p^{\mu }$ is parallel to $x^{\mu }$ and therefore the angular
momentum

\begin{equation}
L^{\mu \nu }=x^{\mu }p^{\nu }-x^{\nu }p^{\mu }  \tag{59}
\end{equation}%
associated with the Lorentz symmetry of (55) should vanish, which is
unlikely result. Thus, if we impose the condition $L^{\mu \nu }\neq 0$ and
the constraints (56)-(58) we find that the signature of $\eta _{\mu \nu }$
should be of at least of the form $\eta _{\mu \nu }=diag(-1,-1,1,...,1).$ In
other words only with two times the constraints (56)-(58) are consistent
with the requirement $L^{\mu \nu }\neq 0$. In principle we can assume that
the number of times is grater than 2 but then one does not have enough
constraints to eliminate all the possible ghosts.

As in section 3 we can rewrite (59) in form

\begin{equation}
L^{\mu \nu }=\frac{1}{2}\varepsilon ^{ab}v_{a}^{\mu }v_{b}^{\nu },  \tag{60}
\end{equation}%
which means that $L^{\mu \nu }$ is a prechirotope. Thus, one of the
conditions for maintaining both the symmetry $SL(2,R)$ and the Lorentz
symmetry in the lagrangian (51) is that the prechirotope $L^{\mu \nu }$ must
be different from zero, in agreement with one of the conditions of the
definition of oriented matroids in terms of chirotopes. Therefore, if our
starting point in the formulation of lagrangian (51) is oriented matroid
theory then the two time physics arises in a natural way.

\bigskip\ 

\noindent \textbf{7.- Final remarks}

\smallskip\ 

Besides the connection between matroid theory and Chern-Simons formalism ,
supergravity, string theory, $p$-branes and Matrix theory found previously,
in this work we have added new links of matroids with different scenarios of
physics such as classical and quantum mechanics and two time physics. All
these physical scenarios are so diverse that one wonders why the matroid
subject has passed unnoticed. This is due, perhaps, to the fact that
oriented matroid theory has evolved putting much emphasis in the equivalence
of various possible axiomatizations. Just to mention some possible
definitions of an oriented matroid besides definition in terms chirotopes
there are equivalent definitions in terms of circuits, vectors and covectors
among others (see Ref. [6] for details). As a result, it turns out that most
of the material in matroid theory is dedicated to existence theorems. Part
of our effort in the present work has been to start the subject with just
one definition and instead of jumping from one definition to another we try
to put the oriented matroid concept, and in particular the chirotope
concept, in such a way that physicists can make some further computations
with such concepts. In a sense, our view is that the chirotope notion may be
the main tool for translating concepts in oriented matroid theory to a
physical setting and vice versa.

It is interesting to mention that even electromagnetism seems to admit a
chirotope construction. In fact, let us write the electromagnetic gauge
potential as [32]

\begin{equation}
A_{\mu }=\varepsilon ^{ab}e_{a}^{i}\partial _{\mu }e_{bi}.  \tag{61}
\end{equation}%
where $e_{a}^{i}$ are two bases vectors in a tangent space of a given
manifold. It turns out that the electromagnetic field strength $F_{\mu \nu
}=\partial _{\mu }A_{\nu }-\partial _{\nu }A_{\mu }$ becomes

\begin{equation}
F_{\mu \nu }=\varepsilon ^{ab}\partial _{\mu }e_{a}^{i}\partial _{\nu
}e_{bi},  \tag{62}
\end{equation}%
We recognize in (62) the typical form a prechirotope (10). The idea can be
generalized to Yang-Mills [32] and gravity using MacDowell-Mansouri
formalism.

As we mentioned an interesting aspect of the oriented matroid theory is that
the concept of duality may be implemented at the quantum level. For
instance, an important theorem in oriented matroid theory assures that

\begin{equation}
(\mathcal{M}_{1}\oplus \mathcal{M}_{2})^{\ast }=\mathcal{M}_{1}^{\ast
}\oplus \mathcal{M}_{2}^{\ast },  \tag{63}
\end{equation}%
where $\mathcal{M}^{\ast }$ denotes the dual matroid and $\mathcal{M}%
_{1}\oplus \mathcal{M}_{2}$ is the direct sum of two oriented matroids $%
\mathcal{M}_{1}$ and $\mathcal{M}_{2}$. If we associate the symbolic actions 
$S_{1}$ $S_{2}$ to the two the matroids $\mathcal{M}_{1}$ and $\mathcal{M}%
_{2}$ respectively, then the corresponding partition functions $Z_{1}(%
\mathcal{M}_{1})$ and $Z_{2}(\mathcal{M}_{2})$ should lead to the symmetry $%
Z=Z^{\ast }$ of the total partition function $Z=Z_{1}Z_{2}.$

Another interesting aspect of duality in oriented matroid theory is that it
may allow an extension in of the Hodge duality. From the observation that
the completely antisymmetric object $\varepsilon _{\mu _{1}...\mu _{d}}$ is
in fact a chirotope associated to the underlaying uniform matroid $U_{n,n}$,
corresponding to the ground set $E=\{1,2,...,n\}$ and bases subset $\mathcal{%
B}=\{\{1,2,...,n\}\}$, it is natural to ask why not to use other chirotopes
to extend the Hodge duality concept? In ref. [23] it was suggested the idea
of the object

\begin{equation}
^{\ddagger }\Sigma ^{\mu _{p+2}...\mu _{r}}=\frac{1}{d!}\chi _{\mu
_{1}...\mu _{p+1}}^{\mu _{p+2}...\mu _{r}}\Sigma ^{\mu _{1}...\mu _{p+1}}, 
\tag{64}
\end{equation}%
where $\Sigma ^{\mu _{1}...\mu _{p+1}}$ is any completely antisymmetric
tensor and $\chi _{\mu _{1}...\mu _{p+1}\mu _{p+2}...\mu _{r}}\equiv \chi
(\mu _{1},..,\mu _{p+1},\mu _{p+2},...,\mu _{r})$ is a chirotope associated
to some oriented matroid of rank $r\geq p+1.$ In [23] the concept $%
^{\ddagger }\mathbf{\Sigma }$ was called dualoid for distinguishing it from
the usual Hodge dual concept

\begin{equation}
^{\ast }\mathbf{\Sigma }^{\mu _{p+2}...\mu _{r}}=\frac{1}{(p+1)!}\varepsilon
_{\mu _{1}...\mu _{p+1}}^{\mu _{p+2}...\mu _{r}}\Sigma ^{\mu _{1}...\mu
_{p+1}}  \tag{65}
\end{equation}%
which is a particular case of (64) when $r=d+1$. It turns out that the
dualiod may be of some interest in $p$-branes theory (see Ref. [23] for
details).

Recently, it was proposed that every physical quantity is a polyvector (see
Ref. [33] and references there in). The polyvectors are completely
antisymmetric objects in a Clifford aggregate. It may be interesting for
further research to investigate whether there is any connection between the
polyvector concept and chirotope concept.

Finally, as it was mentioned the Fano matroid is not orientable. But this
matroid seems to be connected with octonions and therefore with $D=11$
supergravity. Perhaps this suggests to look for a new type of orientability.
Moreover, there are matroids, such as non-Pappus matroid, which are either
realizable and orientable. The natural question is what kind of physical
concepts are associated to these type of matroids. It is tempting to
speculate that there must be physical concepts of pure combinatorial
character in the sense of matroid theory. On the other hand, it has been
proved that matroid bundles have well-defined Stiefel-Whitney classes [8]
and other characteristic classes [11]. In turn, Stiefel-Whitney classes are
closely related to spinning structures. Thus, there must be a
matroid/supersymmetry connection and consequently matroid/M-theory
connection.

\bigskip\ 

\noindent \textbf{Acknowledgments: }I would like to thank M. C. Mar\'{\i}n
for helpful comments.

\bigskip

\smallskip\ \

\end{document}